\documentclass[epj]{svjour}

\usepackage{tabularx}
\usepackage{threeparttable}
\usepackage{xfrac}
\usepackage{microtype}

\newcommand{\isot}[2]{$^{#1}\mathrm{#2}$}
\newcommand{\isott}[2]{\textsuperscript{#1}#2}
\newcommand{\isotn}[2]{$\unboldmath ^{#1}\mathrm{\textmd{#2}}$}

\newcommand{\etKr}{\isot{83}{Kr}}
\newcommand{\etmKr}{\isot{83\mathrm{m}}{Kr}}
\newcommand{\natKr}{\isot{\mathrm{nat}}{Kr}}
\newcommand{\etRb}{\isot{83}{Rb}}

\newcommand{\etKrt}{\isott{83}{Kr}}
\newcommand{\etmKrt}{\isott{83m}{Kr}}
\newcommand{\etRbt}{\isott{83}{Rb}}

\newcommand{\etmKrn}{\isotn{83\mathrm{m}}{Kr}}
\newcommand{\etRbn}{\isotn{83}{Rb}}

\newcommand{\bet}{$\mathrm{\beta{}}$}
\newcommand{\gam}{$\mathrm{\gamma{}}$}

\begin{document}

\title{Gamma-ray energies and intensities observed in decay chain \etRbt{}/\etmKrt{}/\etKrt{}}

\author{M. \v{S}ef\v{c}\'{i}k\inst{1} \and D. V\'{e}nos\inst{1}
 \and O. Lebeda\inst{1} \and C. Noll\inst{2} \and J. R\'{a}li\v{s}\inst{1}}

\institute{Nuclear Physics Institute, Czech Academy of Sciences, \v{R}e\v{z} 130, 25068, Czech Republic \and Helmholtz Institut f\"{u}r Strahlen- und Kernphysik, Universit\"{a}t Bonn, Nussallee 14-16, Bonn, 53115, Germany}

\date{Received: date / Revised version: date}

\mail{sefcik@ujf.cas.cz}

\abstract{Radioactive sources of the monoenergetic low-energy conversion electrons from the decay of isomeric \etmKrn{} are frequently used in the systematic measurements, particularly in the neutrino mass and dark matter experiments. For this purpose, the isomer is obtained by the decay of its parent radionuclide \etRbn{}. In order to get more precise data on the gamma-rays occuring in the \etRbn{}/\etmKrn{} chain, we re-measured the relevant gamma-ray spectra, because the previous measurement took place in 1976. The obtained intensities are in fair agreement with this previous measurement. We have, however, improved the uncertainties by a factor of 4.3, identified a new gamma transition and determined more precisely energies of weaker gamma transitions.
}

\maketitle

\section{Introduction}
\label{intro}

\etmKr{} is formed by the decay of \etRb{} (half-life $86.2\pm{}1$~d) via electron capture (EC). Approximately three quarters of \etRb{} decays result in isomer \etmKr{} ($T_{\sfrac{1}{2}}=$ 1.83~h). It further decays by the cascade of the 9.4 and 32.2~keV nuclear transitions to the \etKr{} ground state. Due to low energy and high multipolarity (E3 for the 32.2~keV transition) of the transitions, the intense conversion electrons are emitted. These monoenergetic electrons are extensively used for the calibration and systematic measurements in the neutrino mass experiments (KATRIN, Project~8)~\cite{KATRIN_2022,Esfahani_2017}, dark matter experiments~\cite{Xiong_2020,Singh_2020} and also in the ALICE and COHERENT projects~\cite{ALICE_2018,Akimov_2021}. In all these experiments the \etRb{} is at first deposited into a suitable substrate, from which the daughter \etmKr{} emanates. The last primary data on the gamma-ray intensities in \etRb{} decay were published several decades ago, see~\cite{Dostrovsky_1964,Vaisala_1976}. The recent compilation and evaluation of the relevant data are available in the Nuclear Data Sheets (NDS)~\cite{McCutchan_2015}. In the frame of our development of the \etmKr{} sources for the neutrino project KATRIN, see~\cite{Suchopar_2018,Lebeda_2022}, we also re-measured the gamma-ray spectra present in the \etRb{} decay.

\section{Measurement}
\label{sec:meas}

Rubidium isotopes were produced at the NPI CAS Řež cyclotron TR-24 in the reactions \natKr{}(p,xn)\isot{83, 84, 86}{Rb} using pressurized gas target. The activity was extracted from the irradiated target by its thorough washout by water. Resulting aqueous solution was concentrated by evaporation and used for the activity deposition into the tungsten furnaces. The furnaces were then delivered to the HISKP in Bonn, where the gamma sources were prepared by the implantation of the separated \etRb{} ions with energy of 8~keV into the 0.5~mm thick Highly Oriented Pyrolytic Graphite (HOPG) substrate. The \etRb{} activity in the sources amounted to about 3 MBq. Another type of the source was prepared in the NPI by evaporation of the rubidium isotopes solution on the 2.5 µm thick mylar foil. For the spectra acquisition, two gamma-ray detectors were used: the Ortec HPGe detector with relative efficiency of 24.1~\% and energy resolution of 1.9~keV at the energy of 1.33~MeV, and low energy Canberra SiLi detector with the diameter and thickness of 10.1 and 5~mm, respectively, and the energy resolution of 180~eV at the gamma-ray energy of 5.9~keV. Both detectors were equipped with a beryllium window. The Canberra spectrometric chains were used for the signal processing: amplifier 2025 and multichannel analyzer Multiport II controlled with the computer software Genie 2000. The ADC gain conversion was set at 8192 and 4096 channels for the HPGe and SiLi detector, respectively. The distance between the detector Be window and the measured source was set to 240 and 45.7~mm for the HPGe and SiLi detector, respectively. In order to reduce the sum peaks of the intense \etRb{} gamma-rays with the energies of 520.4, 529.6 and 552.5~keV with the accompanying strong krypton K X-rays, the nickel foil of 20 µm thickness was applied on the HPGe beryllium window. The measured spectra were analysed with DEIMOS32 software~\cite{Frana_2003}.

The energy and detection efficiency calibrations were performed with use of the standards of \isot{55}{Fe} (type EFX), and \isot{133}{Ba}, \isot{152}{Eu} and \isot{241}{Am} (all three type EG3) provided by the Czech Metrology Institute (CMI). Since the calibration sources are encapsulated in the polymethylmethacrylate (PMMA) and polyethylene, the attenuation of the gamma-rays in these materials was also measured to take it into account in the efficiency calibration. The efficiency of the HPGe detector was calibrated in the low (26--244~keV) and high (244--778~keV) energy regions with the uncertainties of 2.5 and 0.9~\%, respectively. In case of the SiLi detector, the efficiency was determined with the uncertainty of 2~\% in the energy region of 5.9--33~keV.

\begin{figure}
\centering
\includegraphics[width=\columnwidth]{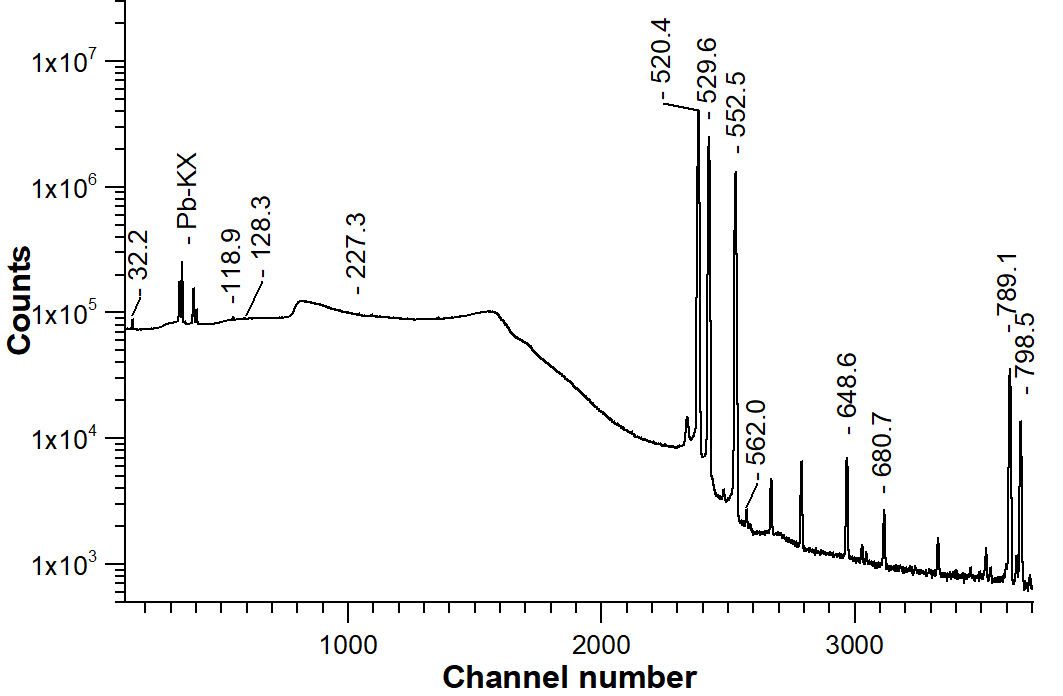}
\caption{Spectrum of \etRb{} acquired with the HPGe detector. The \gam{}-lines which belong to the \etRb{} decay are marked by their energies in~keV. The multiple lines denoted as Pb-KX are due to fluorescence effect in the detector Pb shielding.}
\label{fig:hpge}
\end{figure}

\begin{figure}
\centering
\includegraphics[width=\columnwidth]{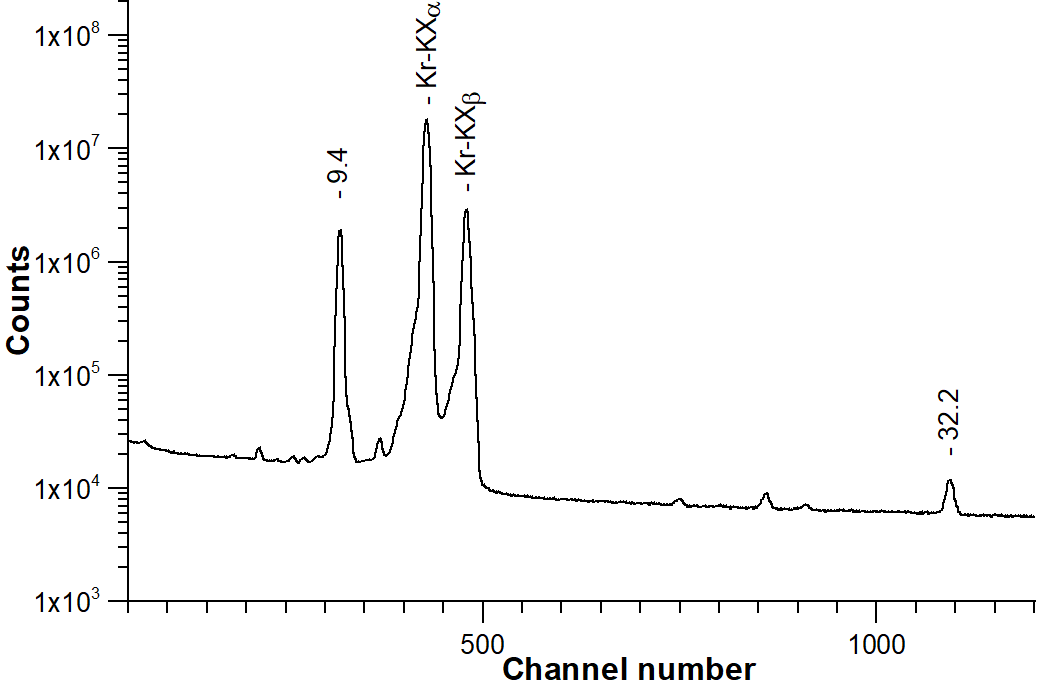}
\caption{The low energy spectrum of \etRb{} acquired with the SiLi detector. Besides the two gamma-rays resulting from the decay of its daughter \etmKr{}, the krypton K X-rays are present.}
\label{fig:sili}
\end{figure}

\begin{figure}
\centering
\includegraphics[width=\columnwidth]{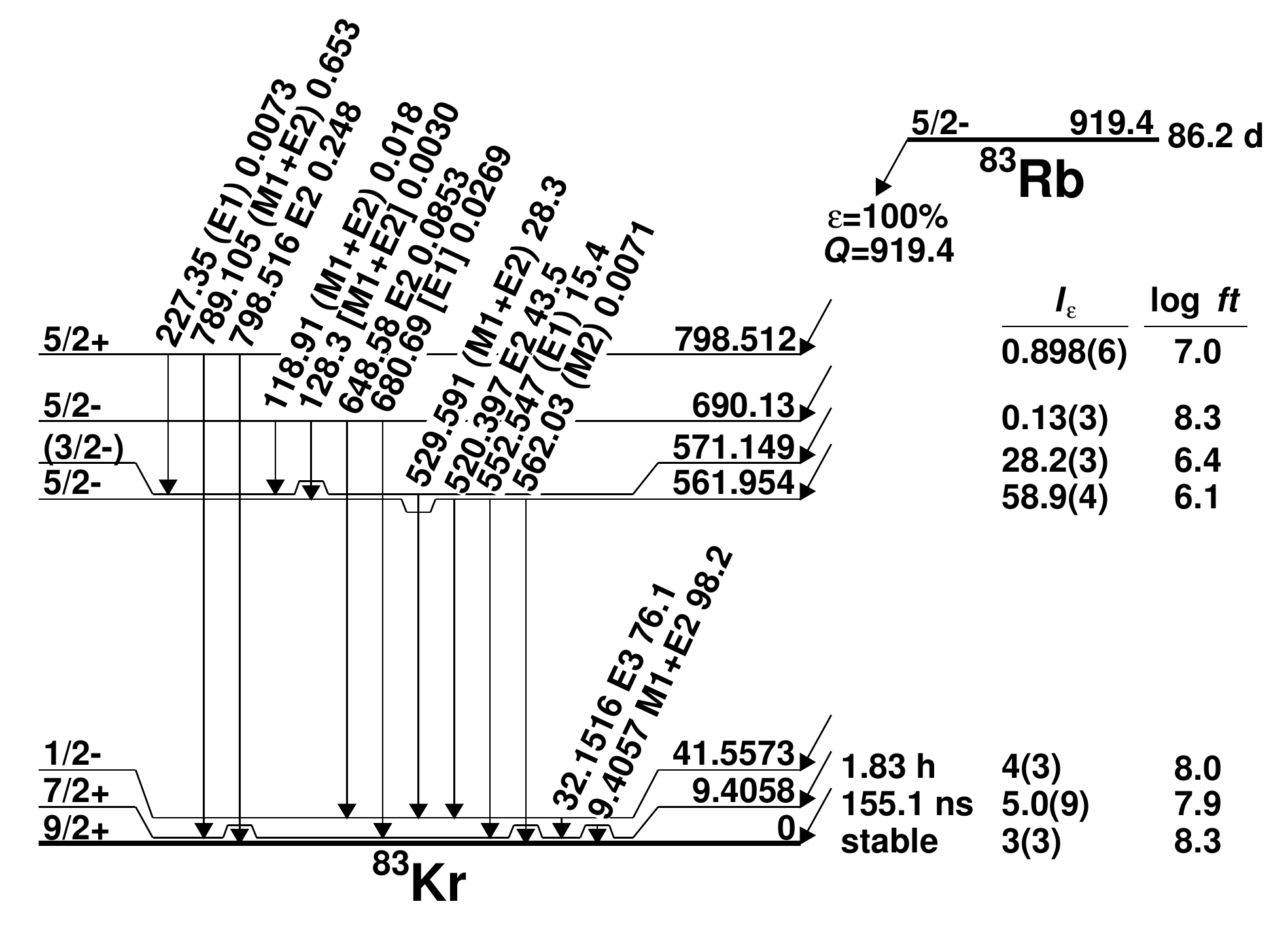}
\caption{Decay scheme of \etRb{}. The spin and parity assignment, half-lives and most multipolarities were adopted from~\cite{McCutchan_2015}. The $\log ft$ values were calculated according to~\cite{Lederer_1978}.}
\label{fig:scheme}
\end{figure}

In Figs.~\ref{fig:hpge} and~\ref{fig:sili}, examples of the \etRb{} gamma-ray spectra measured with the HPGe and SiLi detector, respectively are displayed. The measured gamma-ray energies and intensities are summarized in Tab.~\ref{tab:Rb_gamma}. The gamma-ray energies were determined with the HPGe detector in the special measurements of \etRb{} source together with the \isot{152}{Eu} or \isot{133}{Ba} standards, the gamma-rays of which were used for the calibration of the energy scale. The weak gamma transitions with the energies of 128.3 and 562.03~keV, respectively, were less distinct in the spectra acquired with the standards due to the additional Compton background. That is why the spectra with the sole \etRb{} source were used in their evaluation. For the energy calibration, suitable gamma lines from the background and stronger \etRb{} lines, the energy of which was determined previously by us, were employed. Our energies of the three strongest gamma lines in the \etRb{} decay agree well with the very precise values in~\cite{McCutchan_2015} which were adopted from~\cite{Chang_1993}. The energies of the remaining lines are slightly lower (by 0.1 to 1.0~keV) and were obtained with better precision in comparison with those in~\cite{Vaisala_1976,McCutchan_2015}. We observed all gamma-rays presented in~\cite{McCutchan_2015} except the 237.19~keV one for which the upper limit on its relative intensity of only 0.0011 was estimated. We are not able to observe it due to the presence of the intense 238.632(2)~keV gamma line of the \isot{212}{Pb}~\cite{Auranen_2020} from the \isot{232}{Th} decay chain. In contrast, we observed the gamma line with the energy of 227.35(5)~keV that is missing in~\cite{McCutchan_2015}. The previous NDS review for $\mathrm{A}=83$~\cite{Kocher_1975} listed this transition with the relative intensity of 0.03. The line was clearly visible in the spectra taken with our two different implanted sources. The half-life of this weak line was also determined to be of 90(+21,-12) days, which agrees well with the \etRb{} half-live of 86.2(1) days. This transition also fairly fits into the decay scheme between the nuclear levels 798.5 and 571.1~keV, see Fig.~\ref{fig:scheme}.

\begin{table*}
\centering
\begin{threeparttable}
\caption{The measured \etRb{} gamma-ray energies and relative intensities related to the transition 520.4~keV (100~\%). The total conversion coefficients and the transition intensities in~\% of \etRb{} decays based on the level feeding (see Fig.~\ref{fig:scheme}) are also present. In the last two columns, the energies and intensities from~\cite{McCutchan_2015} are listed for comparison.}
\begin{tabularx}{\textwidth}{ | X | X | X | X | X | X | X | }
\hline
Energy $E_{\mathrm{\gamma{}}}$~(keV) & \gam{}-ray intensity & \hspace{0pt}Multipolarity\hspace{0pt}\tnote{**} & Total conver. coeffic. $\alpha_{tot}$\tnote{**} & Transition inten. (\%) & Energy $E_{\mathrm{\gamma{}}}$~(keV)~\cite{McCutchan_2015} & \gam{}-ray inten.~\cite{McCutchan_2015} \\
\hline
\hline
9.4057(6)\tnote{*} & 13.1(6) & M1+E2 & 16.3(2) & 98(4) & 9.4057(6) & 13(3) \\
\hline
32.1516(5)\tnote{*} & 0.090(3) & E3 & 1950(30) & 76(3) & 32.1516(5) & 0.08(1) \\
\hline
118.91(5) & 0.032(2) & (M1+E2) & 0.3(2) & 0.018(3) & 119.32(9) & 0.032(5) \\
\hline
128.3(1) & 0.006(2) & [M1+E2] & 0.2(2) & 0.0030(9) & 128.6(1) & 0.0030(5) \\
\hline
227.35(5) & 0.017(2) & (E1) & 0.0081(1) & 0.0073(8) & - & - \\
\hline
- & - & & - & & 237.19 & $<0.0011$ \\
\hline
520.397(2) & 100.0(9) & E2 & 0.00283(4) & 43.5(4) & 520.3991(5) & 100(5) \\
\hline
529.591(4) & 65.1(6) & (M1+E2) & 0.00191(3) & 28.3(3) & 529.5945(6) & 66(3) \\
\hline
552.547(4) & 35.5(3) & (E1) & 0.00076(1) & 15.4(1) & 552.5512(7) & 36(2) \\
\hline
562.03(6) & 0.016(1) & (M2) & 0.00499(7) & 0.0071(5) & 562.17(7) & 0.019(2) \\
\hline
648.58(1) & 0.197(2) & E2 & 0.00150(2) & 0.0853(9) & 648.97(5) & 0.19(1) \\
\hline
680.69(3) & 0.062(1) & [E1] & 0.00047(1) & 0.0269(6) & 681.18(7) & 0.07(1) \\
\hline
789.105(3) & 1.51(1) & (M1+E2) & 0.00088(1) & 0.653(6) & 790.15(4) & 1.47(4) \\
\hline
798.516(5) & 0.548(5) & E2 & 0.00086(1) & 0.248(2) & 799.37(5) & 0.53(2) \\
\hline
\end{tabularx}
\begin{tablenotes}
\item [*] The energies for the 9.4 and 32.2 transitions are adopted from~\cite{McCutchan_2015}.
\item [**] The total conversion coefficients were calculated with the software available in~\cite{Kibedi_2008}. In the calculations, the transition multipolarities from~\cite{McCutchan_2015} were used. For the transitions 227.3 and 562.0~keV, not indicated in the reference, the values of E1 and M2 respectively as the most probable were used.
\end{tablenotes}
\label{tab:Rb_gamma}
\end{threeparttable}
\end{table*}

After implantation of the \etRb{}, the amount of the daughter \etmKr{} nuclei in HOPG substrate increases and within several \etmKr{} half-lives, the equilibrium is achieved. The amount of \etmKr{} starts then to decrease practically with the half-life of the parent \etRb{}. The possible emanation of the \etmKr{} out of the substrate may reduce the measured intensities of the 9.4 and 32.2~keV \etmKr{} gamma transitions (the decaying \etmKr{} nuclei find themselves partially outside of the space “visible” by the detector). Therefore we accomplished the measurement of the \etmKr{} retention in the implanted source. For this purpose, the HOPG substrate with the implanted source was placed on the top of the closeable cylindrical chamber. The bottom part of the chamber was equipped with a thin PMMA window enabling the detection of the gamma-rays with the SiLi detector. The chamber design is further described in~\cite{Venos_2014}. Using the 32.2 gamma-ray rates measured with the chamber closed and open at the fixed distance of the HOPG substrate from SiLi detector, the retention of \etmKr{} in the substrate was determined to be 0.974(19), i.e. some emanation occurs.

The relative intensities of the 9.4 and 32.2~keV gamma transitions were corrected for the measured retention value. Our uncertainties of the gamma-ray intensities are on average by a factor of 4.3 smaller in comparison with previously published values. In our \etRb{} decay scheme (Fig.~\ref{fig:scheme}), the feedings of the \etKr{} levels by the electron capture and the $\log ft$ values are listed. An assumption on the feeding of the Kr ground state at a level of 2.5 ± 2.5~\% according to the~\cite{McCutchan_2015} was taken into account. The total intensity of the 32.2~keV transition, representing the number of \etmKr{} nuclei produced per 100 \etRb{} decays, amounts to 76(3)~\%. In contrast to~\cite{McCutchan_2015}, our analysis demonstrated non-zero feeding of 4(3)~\% of the krypton isomeric state from the EC decay.

\section{Conclusion}
\label{sec:con}

We have re-measured the gamma-ray spectra observed in the \etRb{}/\etKr{} decay chain by means of the HPGe and SiLi detectors. The values of the gamma-ray intensities are close to those in the previous paper. Nevertheless, their uncertainties were improved on average by a factor of 4.3. The feeding of the \etKr{} levels from the EC decay with the relevant $\log ft$ values were also determined. We have observed the non-zero feeding of the isomeric state at the level of $4\pm{}3$~\% for the first time. Moreover, the 227.35~keV gamma transition was measured and recommended to be introduced into the \etRb{} decay scheme. The gaseous \etmKr{}, whose monoenergetic electrons are widely used for the systematic physical measurement, is formed in the 76(3)~\% of the \etRb{} decays.

\begin{acknowledgement}

This work was supported by the Ministry of Education, Youth and Sport of the Czech Republic (projects LTT19005 and LM2015056) and the Czech Academy of Sciences.

\end{acknowledgement}


\begin{thebibliography}{}

\bibitem{KATRIN_2022}
KATRIN collaboration, Nat. Phys. \textbf{18}, (2022) 160--166.

\bibitem{Esfahani_2017}
Project~8 collaboration, J. Phys. G: Nucl. Partic. \textbf{44}, (2017) 054004.

\bibitem{Xiong_2020}
W. X. Xiong, M. Y. Guan, C. G. Yang, P. Zhang, J. C. Liu, C. Guo, Y. T. Wei, Y. Y. Gan, Q. Zhao, J. J. Li, Radiat. Detect. Technol. Methods \textbf{4}, (2020) 147--152.

\bibitem{Singh_2020}
A. G. Singh, E. P. Bernard, A. Biekert, E. M. Boulton, S. B. Cahn, N. Destefano, B. N. V. Edwards, M. Gai, M. Horn, N. Larsen, Q. Riffard, B. Tennyson, V. Velan, C. Wahl, D. N. McKinsey, J. Instrum. \textbf{15}, (2020) P01023--P01023.

\bibitem{ALICE_2018}
ALICE collaboration, Nucl. Instrum. Meth. A \textbf{881}, (2018) 88--127.

\bibitem{Akimov_2021}
COHERENT collaboration, J. Instrum. \textbf{16}, (2021) P04002.

\bibitem{Dostrovsky_1964}
I. Dostrovsky, S. Katcoff, A. W. Stoenner, Phys. Rev. \textbf{136}, (1964) B44--B49.

\bibitem{Vaisala_1976}
S. V\"{a}is\"{a}l\"{a}, G. Graeffe, J. Heinonen, A. A. Delucchi, R. A. Meyer, Phys. Rev. C \textbf{13}, (1976) 372--376.

\bibitem{McCutchan_2015}
E. A. McCutchan, Nucl. Data Sheets \textbf{125}, (2015) 201--394.

\bibitem{Suchopar_2018}
M. Suchop\'{a}r, D. V\'{e}nos, O. Dragoun, O. Lebeda, M. Ry\v{s}av\'{y}, J. Sentkerestiov\'{a}, A. \v{S}palek, M. Slez\'{a}k, K. Schl\"{o}sser, M. Sturm, M. Arenz, C. Noll, poster at the 23rd Int. Conf. Neutrino Phys. Astrophys., (2018).

\bibitem{Lebeda_2022}
O. Lebeda, D. Vénos, J. Ráliš, M. Šefčík, O. Dragoun, poster at the 30th Int. Conf. Neutrino Phys. Astrophys., (2022).

\bibitem{Frana_2003}
J. Fr{\'{a}}na, J. Radioanal. Nucl. Chem. \textbf{257}, (2003) 583--587.

\bibitem{Kibedi_2008}
T. Kib{\'{e}}di, T. W. Burrows, M. B. Trzhaskovskaya, P. M. Davidson, C. W. Nestor, Nucl. Instrum. Meth. A \textbf{589}, (2008) 202--229.

\bibitem{Chang_1993}
T. Chang, S. Wang, H. Wang, B. Meng, Nucl. Instrum. Meth. A \textbf{325}, (1993) 196--204.

\bibitem{Auranen_2020}
K. Auranen, E. A. McCutchan, Nucl. Data Sheets \textbf{168}, (2020) 117--267.

\bibitem{Kocher_1975}
D. C. Kocher, Nucl. Data Sheets \textbf{15}, (1975) 169--201.

\bibitem{Venos_2014}
D. V{\'{e}}nos, M. Slez{\'{a}}k, O. Dragoun, A. Inoyatov, O. Lebeda, Z. Pulec, J. Sentkerestiov{\'{a}}, A. {\v{S}}palek, J. Instrum. \textbf{9}, (2014) P12010.

\bibitem{Lederer_1978}
C. M. Lederer, V. S. Shirley, \textit{Table of isotopes} (John Wiley \& Sons, Inc., Hoboken, New Jersey, USA 1978) Appendix V, 19.

\end{thebibliography}
\end{document}